\documentclass[doublecol]{epl2} 

\title{Superdiffusive, heterogeneous, and collective
particle motion near the fluid-solid transition
in athermal disordered materials}
\shorttitle{Collective particle motion near fluid-solid transition}

\author{Claus Heussinger\inst{1} \and Ludovic Berthier \inst{2} \and Jean-Louis
  Barrat\inst{1}}

\institute{ \inst{1} Universit\'e de Lyon; Univ. Lyon I, Laboratoire de Physique
  de la Mati\`ere
  Condens\'ee et Nanostructures; UMR CNRS 5586, 69622 Villeurbanne, France \\
  \inst{2} Laboratoire des Collo{\"\i}des, Verres et Nanomat{\'e}riaux, UMR CNRS
  5587, Universit{\'e} Montpellier 2, 34095 Montpellier, France }


\abstract{ We use computer simulations to study the microscopic dynamics of an
  athermal assembly of soft particles near the fluid-to-solid, jamming
  transition. Borrowing tools developed to study dynamic heterogeneity near
  glass transitions, we discover a number of original signatures of the jamming
  transition at the particle scale.  We observe superdiffusive, spatially
  heterogeneous, and collective particle motion over a characteristic scale
  which displays a surprising non-monotonic behavior across the transition.  In
  the solid phase, the dynamics is an intermittent succession of elastic
  deformations and plastic relaxations, which are both characterized by
  scale-free spatial correlations and system size dependent dynamic
  susceptibilities.  Our results show that dynamic heterogeneities in dense
  athermal systems and glass-formers are very different, and shed light on
  recent experimental reports of `anomalous' dynamical behavior near the jamming
  transition of granular and colloidal assemblies.  }

\usepackage{epsfig,amssymb,amsfonts,amsmath}

\newcommand{\be}{\begin{equation}}
\newcommand{\ee}{\end{equation}}

\begin{document}

\maketitle

Many materials, from emulsions and suspensions to foams and granular materials
are dense assemblies of non-Brownian particles~\cite{larson}. Since thermal
energy is irrelevant, the dynamics of these systems must be studied in driven
non-equilibrium conditions. Together with the driving mechanism, a second
important control parameter is the volume fraction of the particles, $\phi$.
These materials undergo a fluid-to-solid `jamming' transition as $\phi$
increases beyond some critical density $\phi_c$~\cite{jammingbook}.  


The properties of this jamming transition have received considerable interest in
recent years, and much progress was made through the analysis of idealized
theoretical models, such as soft, frictionless, repulsive particles as studied
below~\cite{martinreview}.  Structural and mechanical properties of systems on
both sides of the jamming transition have been analyzed~\cite{torquato,ohern},
and a number of remarkable features emerged, such as algebraic scaling of linear
mechanical response~\cite{ohernpre03}, or the development of nontrivial
timescales or lengthscales characterizing the macroscopic behavior of the
system~\cite{wyart,teitel,hatano}.

With the developments of new experimental techniques (confocal microscopy,
original light scattering techniques, etc.), the dynamics of systems near the
jamming transition can now be resolved at the particle scale. Very recently, a
number of `anomalous' or `unexpected' dynamic behaviors were reported in
granular and colloidal assemblies: superdiffusive particle
motion~\cite{luca,lechenaut}, nonmonotonic variations of characteristic scales
across the transition~\cite{lechenaut,candelier,trappe,tapioca}, anomalous drop
of dynamic correlations upon compression~\cite{trappe,ballesta}.  Thus, it
appears timely to investigate, at the fundamental level of single particle
trajectories, the signatures of the jamming transition in idealized theoretical
models. Our work reveals original signatures of the jamming transition in both
fluid and solid phases which, we believe, shed light on recent experimental
findings. We also establish a connection between single particle dynamics and
collective particle motion, which allows us to develop an intuitive and
appealing picture of jamming as the consequence of the diverging size of rigid
particle clusters.


We study a bidimensional 50:50 binary mixture of $N$ particles of diameter ratio
1.4 with harmonic repulsion~\cite{durian}, $V(r_{ij}) = k
(r_{ij}-\sigma_{ij})^2$, where $r_{ij}$ is the distance between the centers of
particle $i$ and $j$, $\sigma_{ij} = (\sigma_i+ \sigma_j)/2$, and $\sigma_i$ is
the diameter of particle $i$.  Particles only interact when they overlap,
$V(r_{ij} > \sigma_{ij})=0$. We work in the zero temperature limit, so that all
the dynamical processes in the system are induced by an external driving. We use
a simple shear flow, and perform quasi-static shear simulations, as described
before~\cite{claus}.  Very small shear strains are applied, followed by a
minimization of the potential energy.  At each step, particles are first
affinely displaced along the $y$-axis by an amount $\delta y = \gamma_0 x$ with
$\gamma_0 = 5 \times 10^{-5}$.  Appropriate Lees-Edwards periodic boundary
conditions are used.  During the subsequent energy minimization, additional
nonaffine displacements occur in both directions.  In the dilute limit, these
nonaffine displacements are absent, and their presence directly reveals the
influence of interactions and interparticle correlations. In the following, we
focus on the purely nonaffine displacements occurring along the $x$-direction,
i.e. transverse to the flow.  We use system sizes $N=900$, 1600 and 2500 to
detect finite size effects. The unique control parameter is the volume fraction,
$\phi$. Below a critical value, $\phi_c \simeq 0.843$, which is well defined and
sharp in the thermodynamic limit, the system flows with no resistance, while a
yield stress grows continuously from 0, when $\phi$ increases above
$\phi_c$~\cite{heu}.  This corresponds to the jamming transition for the present
system and driving conditions.

\begin{figure}
\psfig{file=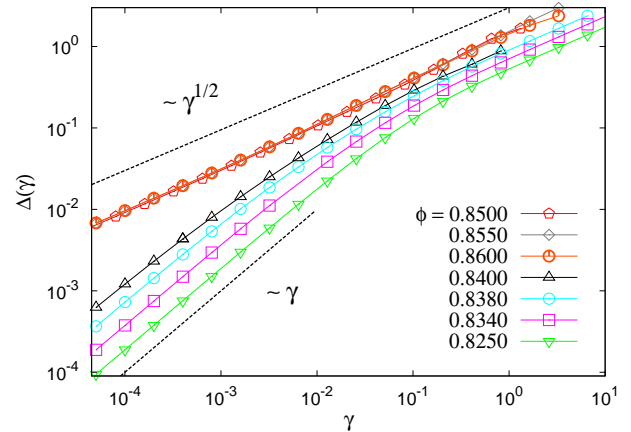,height=8.5cm,angle=-90}
\caption{\label{fig:msqdisp}
Nonaffine displacement $\Delta(\gamma)$ for various volume fractions $\phi$.
For $\phi>\phi_c$, diffusive motion is observed at all $\gamma$.
In contrast, a superdiffusive regime develops
for $\phi<\phi_c$, with a particle `velocity'
 $\ell_\Delta(\phi) = \Delta(\gamma) / \gamma$ which
increases when approaching $\phi_c$.}
\end{figure}

The simplest quantity which is measured from particle displacements is the root
mean-squared displacement, $\Delta (\gamma) = \left\langle \frac{1}{N}
  \sum_{i=1}^N [x_i(\gamma) - x_i(0)]^2 \right\rangle^{1/2},$ where
$x_i(\gamma)-x_i(0)$ is the transverse displacement of particle $i$ after a
total strain $\gamma$ is applied.  In this expression, the brackets indicate
averages taken over independent initial conditions all chosen in the driven
stationary state at a particular density.  We present the evolution of
$\Delta(\gamma)$ at various volume fractions across $\phi_c$ in
Fig.~\ref{fig:msqdisp}.  The displacements in the yield-stress flow regime above
$\phi_c$ are diffusive at all strains, $\Delta \sim \sqrt{\gamma}$.
This is a robust signal of the quasi-static dynamics in yield-stress materials.
It can be traced back to the accumulation of plastic rearrangements spanning the
entire system~\cite{lemaitreCaroliPRE2007,tanguyEPJE2006}. The important
observation here is that the diffusion constant depends  only weakly  on density
and distance to $\phi_c$.

The regime below $\phi_c$ is more interesting. At any given volume fraction, particles
move in a superdiffusive, ballistic manner at short strain, $\Delta \sim
\gamma$, before crossing over to diffusive motion at large strain. Since we work
at zero temperature, this ballistic regime is completely unrelated to the trivial
short-time ballistic displacements in particle systems with classical Newtonian dynamics.

Superdiffusive behavior can be interpreted from the observation that below the
jamming transition, the network of particle contacts is not sufficient to insure
mechanical stability, and there exists a large number of zero-frequency modes
allowing particle displacements at no energy cost~\cite{alexander}. The actual
displacement field is a superposition of these modes, which can thus persist
until the particle configuration has evolved sufficiently to decorrelate the
mode spectrum. Persistence of the modes, followed by their decorrelation explain
superdiffusive and diffusive motion, respectively.

Remarkably, Fig.~\ref{fig:msqdisp} shows that particles move faster when $\phi$
is increased towards $\phi_c$ in both ballistic and diffusive regimes. In
particular, the short-time particle `velocities' {$\ell_\Delta(\phi)\equiv
\lim_{\gamma\to 0 }\Delta(\gamma)/\gamma$} increase with $\phi$. This is a surprising finding
because compression towards jamming usually yields slower
dynamics~\cite{lechenaut,ballesta}.  Note that $\ell_\Delta$ is a displacement
per unit of strain and thus has the dimension of length. Thus an increasing
particle velocity in fact implies an increasing length scale. Indeed, we will
argue below that superdiffusion reflects the concerted motion of solid-like
clusters of particles correlated over a length scale comparable to
$\ell_\Delta(\phi)$, and which grows towards $\phi_c$.

The displacements $\Delta$ and the associated length scale $\ell_\Delta(\phi)$
measure how much particles have to move in addition to the affine displacements,
in order to accomodate the imposed shear flow. At $\phi=0.840$, for example,
this additional displacement is about ten times larger than the affine
contribution itself. This means, that close to $\phi_c$ the system is in a
highly `fragile' state and small changes in the boundary or loading conditions
lead to large-scale motions~\cite{fragile}.

Interestingly, superdiffusion was recently used to identify the location of the
jamming transition in a bidimensional granular system~\cite{lechenaut}.  Due to
the `random agitation' driving mechanism in that experiment, superdiffusion is
observed on a modest time window, and only very close to $\phi_c$.  This is
consistent with our finding that superdiffusion is indeed most pronounced near
the transition.


A second, often studied, correlation function to quantify
single particle dynamics is the `overlap' function~\cite{lechenaut}
\be
Q(a,\gamma) = \left\langle \frac{1}{N} \sum_{i=1}^N
\exp\left( -
\frac{[ x_i(\gamma) - x_i(0) ]^2}{2 a^2} \right)
\right\rangle.
\label{eq:Q_definition}
\ee
The overlap $Q(a,\gamma)$ goes from 1 to 0 as typical
particle displacements get larger than the probing
length scale $a$. It is thus very similar to a self-intermediate
scattering function in liquids, with $a$ playing the role of an inverse
wavevector, and $\gamma$ the role of time.

\begin{figure}
\psfig{file=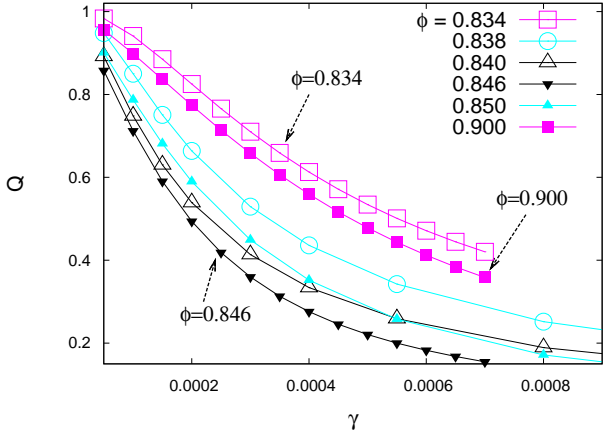,height=6.0cm,angle=-0}
\caption{\label{fig:q.unscaled}
The average overlap $Q(a,\gamma)$ for $a=0.001$ and different
volume fractions across the jamming transition.
The relaxation becomes faster when $\phi$ increases
towards $\phi_c$, but slows down when $\phi$ increases
above $\phi_c$. The fluid and solid phases are thus readily
distinguished by their distinct behavior upon compression.}
\end{figure}

The behavior of $Q(a,\gamma)$ is presented in Fig.~\ref{fig:q.unscaled} for a
fixed $a=0.001$ (qualitatively similar results are obtained for different
choices of $a$) and various volume fractions across $\phi_c$.  For $\phi <
\phi_c$, we find that the overlap function decays faster when $\phi$ increases.
This is consistent with the above observation that displacements get larger
closer to $\phi_c$. A qualitatively different behaviour is found above the
transition, where the overlap decays more slowly when $\phi$ increases. This is
in striking contrast with the behavior of the mean squared displacement in
Fig.~\ref{fig:msqdisp} which showed no such variations with $\phi$.

The data in Fig.~\ref{fig:q.unscaled} thus imply the existence of an unexpected
nonmonotonic variation of the dynamical behaviour, which is absent in the root
mean-squared displacement, Fig.~\ref{fig:msqdisp}. To quantify these differences
we define the analog $\ell_Q(\phi)$ of the displacement scale,
$\ell_\Delta(\phi)$, discussed above. For a given strain $\gamma$ we measure
the value of $a$ such that 
\be Q \left( a \equiv \gamma\ell_Q(\phi), \gamma \right) = 0.5\,.
\label{05}
\ee The value 0.5 is arbitrary and we find similar results with 0.3 and 0.7.
{For small strains within the ballistic regime this definition insures
  that $\ell_Q$ is independent of strain $\gamma$ and reflects a short-time
  velocity, as does its analog $\ell_\Delta$. For larger strains a crossover to
  diffusive behavior sets in similar as for the root mean-square displacement
  $\Delta$ (Fig.~\ref{fig:msqdisp}). In the following, we want to avoid this
  crossover and thus choose $\gamma$ as small as possible.}

\begin{figure}
\psfig{file=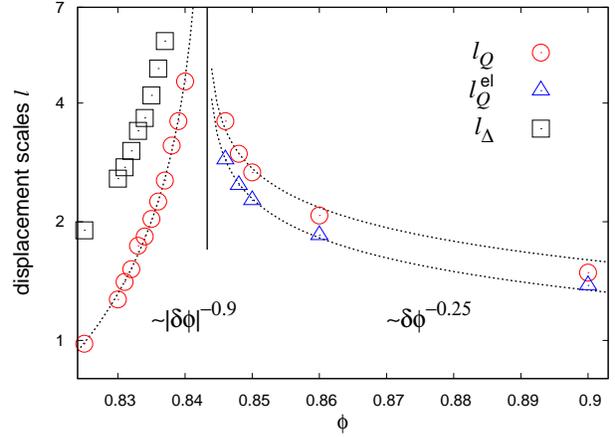,height=6.0cm,angle=0}
\caption{\label{fig:vhmax.above.below} Relaxation length scale $\ell_Q(\phi)$
  quantifying the decay of the overlap $Q(a,\gamma)$ for {a single
    strain-step $\gamma=\gamma_0$}, from Eq.~(\ref{05}) ($\ell_Q^{\rm el}$ is
  defined from purely elastic deformations above jamming).  It is a nonmonotonic
  function, with a maximum at the jamming transition, mirroring the behavior of
  the overlap in Fig.~\ref{fig:q.unscaled}.  It obeys power law behavior with
  different exponents on both sides of the transition.  In the fluid phase,
  $\ell_Q$ behaves similarly to the `velocity' $\ell_\Delta$ defined from
  superdiffusive behaviour. }
\end{figure}

In Fig.~\ref{fig:vhmax.above.below} we report the behaviour of $\ell_Q(\phi)$ as
defined from Eq.~(\ref{05}), which clearly reflects the nonmonotonic behaviour
of the overlap observed in Fig.~\ref{fig:q.unscaled}. This result shows that a
remarkably simple statistical analysis of single particle displacements in
athermal systems very easily reveals the existence of, and quite accurately
locates, the fluid-to-solid jamming transition. This measurement requires no
finite size scaling, or other involved or indirect statistical
analysis~\cite{lechenaut,candelier,ballesta}.

We have discussed two  signatures of the jamming transition from the
behavior of $\Delta$ (superdiffusion) and $Q(a,\gamma)$ (nonmonotonic 
relaxation as a function of density).  
It should come as a third surprise that both dynamic quantities
apparently provide distinct informations, in particular for $\phi > \phi_c$,
although they both aim at quantifying dynamics at the particle scale. In fact,
such discrepancies are well-known and well-studied in the field of the glass
transition where several similar `decoupling' phenomena have been
studied~\cite{ediger}. Indeed, if the particle dynamics were 
a Gaussian process,
the information content of $\Delta$ and $Q$ would be mathematically equivalent.
Their different behavior thus suggests that the distribution of single particle
displacements is strongly non-Gaussian~\cite{pinaki}.

\begin{figure}
\psfig{file=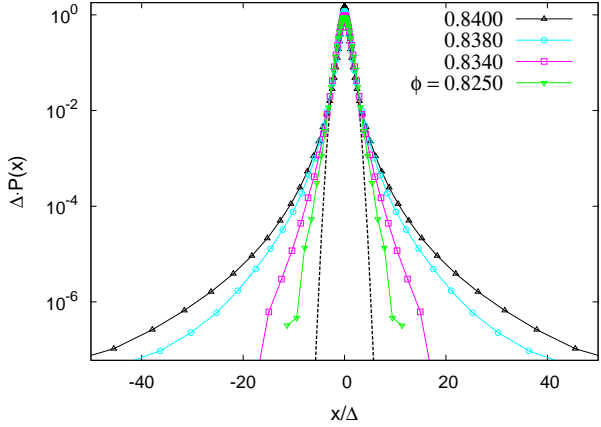,height=8.5cm,angle=-90}
\caption{Displacement distribution after a single strain step,
  $\gamma=\gamma_0$, for different volume fractions $\phi < \phi_c$,
rescaled such that a
  Gaussian process (dashed line)
has unit variance. Increasingly heterogeneous distributions 
of particle displacements are found when approaching $\phi_c$,
with the development of algebraic tails yielding a diverging 
kurtosis at $\phi_c$.
\label{fig:vanHove}}
\end{figure}

To substantiate this claim we report the evolution of the van Hove distribution,
$P (x,\gamma) = \left\langle \frac{1}{N} \sum_i\delta \left( x - [x_i(\gamma) -
    x_i(0)] \right)\right\rangle $, as a function of volume fraction below
$\phi_c$ (Fig.~\ref{fig:vanHove}).
Above $\phi_c$ the shape of the van Hove function has a form well-known from
recent work on sheared glasses~\cite{lemaitreCaroliPRE2007,tanguyEPJE2006};
therefore, we do not show our data here. Briefly, the distribution is `bimodal',
with a first component at small $x$ resulting from displacements during
reversible elastic branches, and a second component at larger $x$ due to
irreversible plastic events. This latter contribution completely dominates the
second moment of the distribution, and thus controls the diffusive behavior of
$\Delta(\gamma)$ in Fig.~\ref{fig:msqdisp}.

The decomposition between elastic and plastic branches above $\phi_c$ suggests a
separate analysis of these two types of dynamics. We have computed $Q(a,\gamma)$
along elastic branches only, and defined $\ell^{\rm el}_Q(\phi)$, the
`elastic' analog of the displacement scale $\ell_Q(\phi)$ for the full
overlap; it is shown in Fig.~\ref{fig:vhmax.above.below}.  Both $\ell_Q$ and
$\ell^{\rm eq}_Q$ are very close, confirming that $Q(a,\gamma)$ is not very
sensitive to plastic rearrangements, in contrast to $\Delta(\gamma)$.
The scaling of the typical elastic response above $\phi_c$ has recently been
discussed in great detail~\cite{martin2,wyart2,martin3}, and we simply quote the
result: $\ell_Q^{\rm el} \sim (\phi_c - \phi)^{-1/4} \sim \ell_Q$, in good agreement
with our numerical results.  

Below $\phi_c$ the distributions remain highly non-Gaussian
(Fig.~\ref{fig:vanHove}), even though plastic events are absent. Deviations from
a Gaussian shape become stronger as $\phi$ increases towards $\phi_c$. The
distribution develops polynomial tails, and it is thus not surprising that
different averages taken over such distributions provide quantitatively distinct
results.
In Fig.~\ref{fig:vhmax.above.below}, we show that $\ell_Q \sim (\phi_c -
\phi)^{-0.9}$, for $\phi < \phi_c$. In contrast, we find that the scale
$\ell_\Delta$, defined in Fig.~\ref{fig:msqdisp}, scales as $\ell_\Delta \sim
(\phi_c - \phi)^{-1.1}$ with an exponent slightly distinct from the one of
$\ell_Q$.  Hence, despite the appeal of the algebraic `scaling' relations
reported in Fig.~\ref{fig:vhmax.above.below}, our particle distributions
$P(x,\gamma)$ do not follow simple scaling forms, even along elastic branches,
in apparent contradiction with recent claims~\cite{martin3}.  Single particle
dynamics is not `universal' near the jamming transition and it is somewhat
ambiguous to define `typical' displacements from such broad distributions.

It is tempting to speculate that the polynomial tails which develop near but
below $\phi_c$ in Fig.~\ref{fig:vanHove} are in fact the `precursors' of plastic
events taking place above $\phi_c$.
The tails in fact become so broad that the kurtosis of the distribution (also
called `non-Gaussian parameter') actually diverges at small $\gamma$ when
$\phi_c$ is approached from below (data not shown). This implies that the
kurtosis could be used as a second, simple statistical indicator of the
underlying jamming transition. In glass-formers, the non-Gaussian parameter
increases as the glass transition is approached, but much more
modestly~\cite{walter}, while van hove distributions develop
exponential rather than algebraic tails~\cite{pinaki}.

\begin{figure}
\includegraphics[width=\columnwidth]{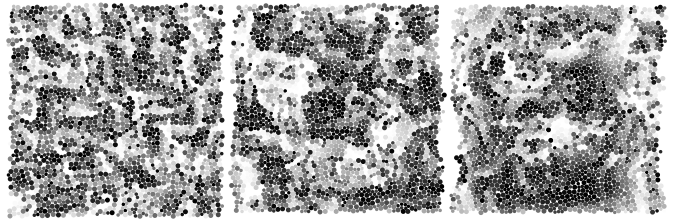}
\caption{\label{fig:mobility_displ} Grey-scale plot of the spatial fluctuations
  of the overlap in Eq.~(\ref{eq:Q_definition}) (dark=`immobile',
  light=`mobile') for $\phi = 0.825 <\phi_c$ (left), $\phi_c = 0.840$ (middle)
  and $\phi = 0.85 > \phi_c$ (right). Strain {$\gamma=\gamma_0$} and probing
  length scale $a$ chosen such that $Q \approx 0.5$.  Spatial correlations are
  increasing with $\phi$ towards $\phi_c$, but stay comparable to system size
  above $\phi_c$ when the system is in the solid phase.}
\end{figure}

Let us now turn to the discussion of particle correlations. This will allow us
to make concrete the notion of `rigid' clusters that we have alluded to above.
To this end, we first discuss the images shown in Fig.~\ref{fig:mobility_displ},
where the spatial fluctuations of the overlap are shown. From this figure, we
can readily identify dynamically correlated clusters of particles, that clearly
grow in size upon approaching $\phi_c$ from below.  These images are direct
evidence that dynamics becomes more collective as $\phi_c$ is approached.  Above
$\phi_c$ the system is now solid, and responds as an elastic body.
Correspondingly, mobility fluctuations are correlated on a length scale
comparable to the system size, the only visible effect of the density being that
the snapshot looks less `disordered' at the largest $\phi$. Very similar observations
were recently made experimentally in a soft granular material~\cite{tapioca}.

\begin{figure}
\psfig{file=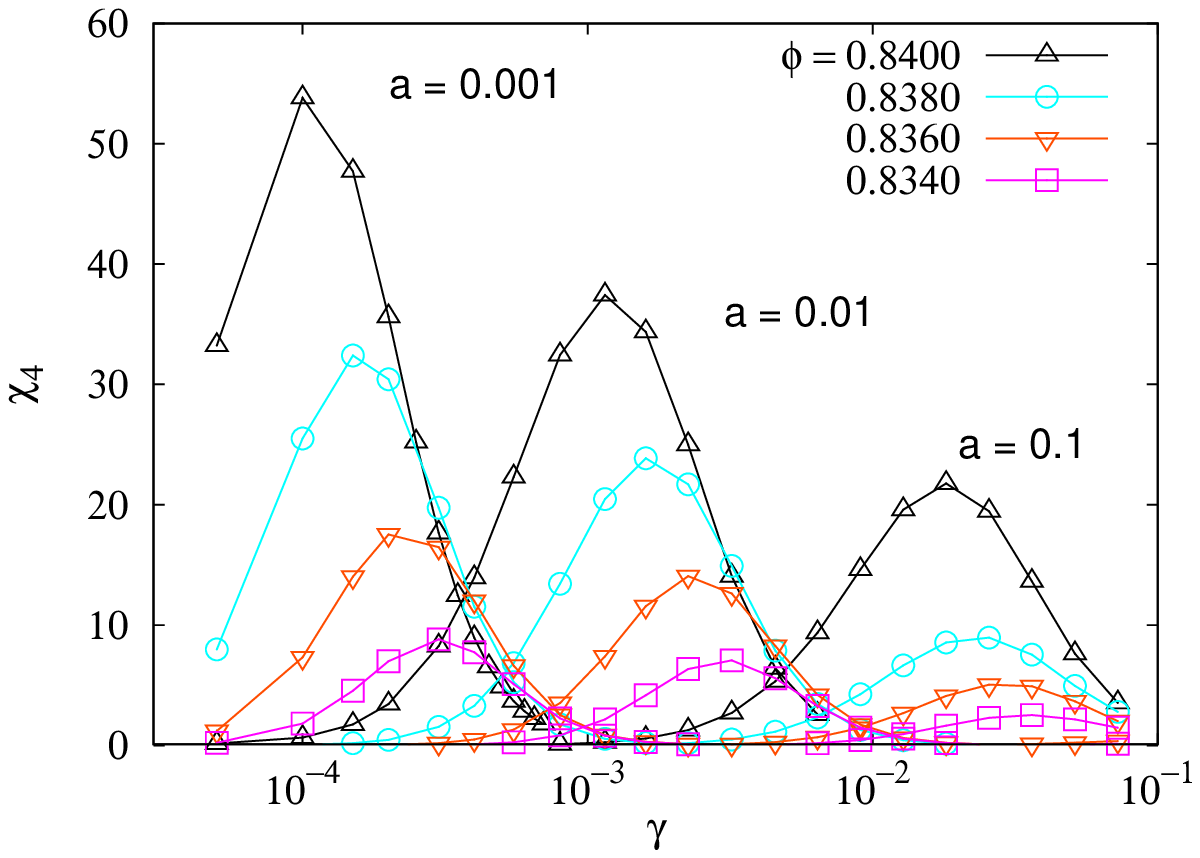,height=6cm,angle=0} 
\psfig{file=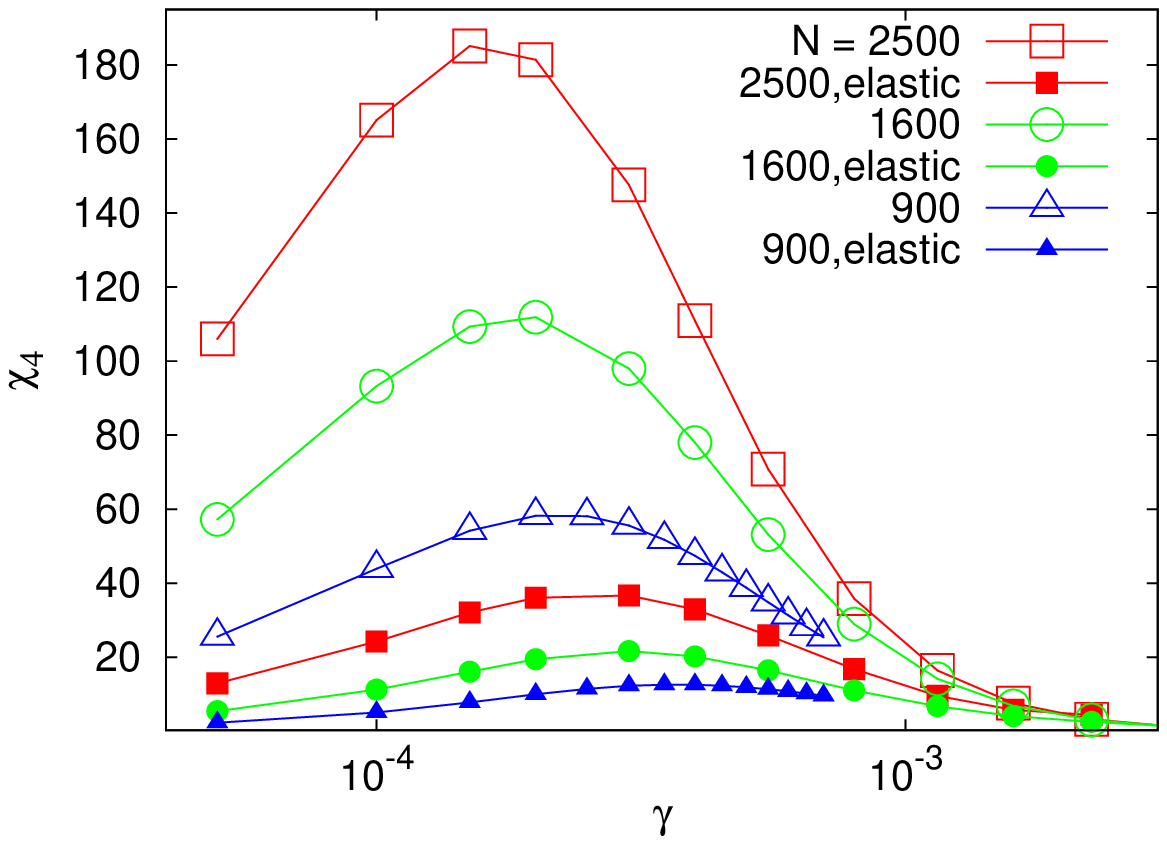,height=6cm,angle=0}
\caption{\label{fig:chi4.below}
Dynamical susceptibility $\chi_4$ as function of strain $\gamma$
for: (Top) different volume-fractions $\phi<\phi_c$ and probing length scale
$a$. (Bottom) a single $\phi = 0.85 > \phi_c$ and different system sizes.
The dynamic susceptibility increases when $\phi$ increases
towards $\phi_c$ reflecting increasingly collective dynamics.
Above $\phi_c$ dynamic correlations are system size dependent, even
when only elastic branches are considered.}
\end{figure}

To quantify these qualitative observations, we study the dynamical
susceptibility $\chi_4$, defined as the variance of statistical fluctuations of
the overlap $Q$~\cite{franz}, $\chi_4(a,\gamma) = N \left(\left\langle
    \hat{Q}(a,\gamma)^2 \right\rangle - Q(a,\gamma)^2 \right)$, where $\hat{Q}$
represents the instantaneous contribution to the average $Q$.  Our results are
summarized in Fig.~\ref{fig:chi4.below}.  In all cases, $\chi_4(a,\gamma)$
displays a well-defined maximum when shown as a function of $\gamma$, which
shifts towards lower strain when $\phi$ increases. This maximum identifies a
strain for which fluctuations of the overlap are maximal, and is slaved to the
typical strain over which $Q$ decays in Fig.~\ref{fig:q.unscaled}, as commonly
found in studies of dynamic heterogeneity~\cite{glotzer,toni}.  As is
well-known, the height of this maximum is a good estimate of the number of
particles relaxing in a correlated manner~\cite{toni}, which justifies
the fundamental importance of $\chi_4$. The increase of the height of the peak
of the dynamic susceptibility when $\phi$ increases towards $\phi_c$ in
Fig.~\ref{fig:chi4.below} thus quantitatively confirms the visual impression
given by the snapshots: the dynamics is increasingly collective when $\phi$
increases towards $\phi_c$. As shown in Fig.~\ref {fig:chi4.below}, this
conclusion does not depend on our choice of a probing length scale $a$.

\begin{figure}
\psfig{file=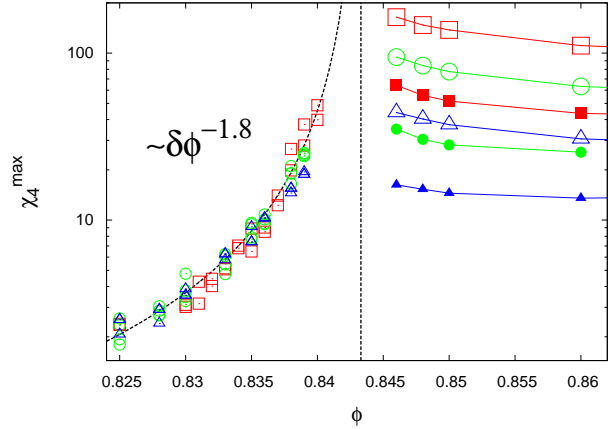,width=8.5cm,angle=0}
\caption{\label{fig:chi4.maximum} Evolution of the maximum of the dynamic
  susceptibility $\chi_4$ across the jamming transition ({$\gamma=\gamma_0$};
  symbols as in Fig.~\ref{fig:chi4.below}). It diverges algebraically as $\phi$
  increases towards $\phi_c$ and is system size independent.  Above $\phi_c$,
  the maximum is proportional to the system size, with an amplitude which decays
  slowly upon compression, both for plastic and elastic dynamics.}
\end{figure}

To quantify the possible divergence of the spatial correlation of the dynamics,
we analyze the data as above for the overlap. For a given strain $\gamma$, we
look for $a$ such that $\chi_4$ is maximum (it corresponds roughly to
$\ell_Q(\gamma)$ defined in Eq.~(\ref{05})) and measure its height.  We report
the evolution of this maximum as a function of volume fraction in
Fig.~\ref{fig:chi4.maximum}.  
The data for three different system sizes superimpose below $\phi_c$, and we can
satisfactorily describe the $\phi$-dependence as: $\chi_4 \sim (\phi_c -
\phi)^{-1.8}$.  Comparing this apparent divergence with the much more modest
increase of dynamic correlations in glass-forming liquids close to the glass
transition~\cite{science05}, emphasizes once more the qualitative differences
between both types of transitions.

Assuming that correlated domains have a compact geometry, this finding
immediately translates into a genuine dynamic correlation length scale, $\xi_4$,
which grows as $\xi_4 = \sqrt{\chi_4} \sim (\phi_c - \phi)^{-0.9}$.  Strikingly,
we found a similar power law divergence for both $\ell_Q$ and $\ell_\Delta$ in
Figs.~\ref{fig:msqdisp} and \ref{fig:vhmax.above.below}. The exponents are
furthermore close to the one obtained in previous work \cite{claus} using the
spatial dependence of the displacement correlation function.

These findings call for a simple relation between single-particle displacements,
as characterized by $\ell_Q$ or $\ell_\Delta$, and particle correlations, as
given by $\xi_4$ or by displacement correlations.  The picture that we propose
assumes particles to form `temporarily rigid' clusters of the size of the
correlation length $\xi_4$. Driven by the shear-strain $\gamma$ these clusters
move or rotate in a solid-like manner over distances $\Delta_{\rm cluster} =
\xi_4\cdot\gamma$.  This motion shows up as pronounced ballistic regime in the
mean-squared displacement or in the decay of the overlap function $Q$ and allows
to make the identification, $\ell_Q\simeq \xi_4$.  The cross-over to particle
diffusion then corresponds to a typical cluster lifetime.  On longer strains,
clusters break up and lose their identity. This process is evident in the
reduction of the dynamical correlation length as measured by the decreasing
amplitude of $\chi_4$ with $\gamma$ in Fig.~\ref{fig:chi4.below}.


Moving to $\phi > \phi_c$ in Fig.~\ref{fig:chi4.below}, we observe that $\chi_4$
still exhibits a maximum at a given strain $\gamma$, which again tracks the
relaxation of $Q$.  Remarkably, we now find that the height of this peak
strongly depends on the system size, and increases roughly linearly with $N$,
see Fig.~\ref{fig:chi4.below}.  This suggests that spatial correlations saturate
to the system size at $\phi_c$ and remain scale-free above $\phi_c$ in the
entire solid phase\cite{heu}. This is in agreement with studies of the mechanical response
of disordered solids, which showed evidence of plastic rearrangements spanning
the entire system~\cite{lemaitre,tanguyEPJE2006}.  Additionally, we show in
Figs.~\ref{fig:chi4.below} and \ref{fig:chi4.maximum} that the dynamic
susceptibility measured along purely elastic branches (discarding plastic
events), is linearly growing with $N$ as well. 
Indeed, the elastic response of a disordered solid
is also scale-free, as shown theoretically~\cite{lubensky,picard} and
numerically~\cite{maloney,barrat}.

For a given system size, $\chi_4$ grows with $\phi$ below $\phi_c$, and becomes
proportional to $N$ above $\phi_c$. However, the amplitude of $\chi_4$ decreases
slowly with $\phi$ above $\phi_c$. The net result is that $\chi_4$, for a fixed
system size $N$, has a striking nonmonotonic behaviour with $\phi$, and presents
an absolute maximum at $\phi_c$, which adds to our list of microscopic
signatures of the jamming transition.

A similar maximum of $\chi_4$ has recently been reported both for
colloidal~\cite{trappe,ballesta} and granular~\cite{lechenaut} assemblies, and
given two distinct interpretations.  It was associated to a nonmonotonic dynamic
correlation length scale $\xi_4$ diverging at $\phi_c$ in Ref.~\cite{lechenaut}.
Alternatively, it was attributed to a nonmonotonic strength of essentially
scale-free spatial correlations in Refs.~\cite{trappe,ballesta}.  Our results
clearly favor the second interpretation to explain the behavior above $\phi_c$.
The dynamic susceptibility $\chi_4$ above $\phi_c$ is system size dependent with
scale-free spatial dynamic fluctuations (Fig.~\ref{fig:chi4.maximum}).

This finding suggests that the length scale identified from the low-frequency
part of the vibrational spectrum, and much studied
recently~\cite{wyart,martin3,barrat} does not directly influence the dynamics at
the particle scale beyond the linear regime where it belongs.  We suggest,
however, that it indirectly influences the behaviour of $\chi_4$, and leads to
the decaying strength of the correlations as observed in
Fig.~\ref{fig:chi4.maximum} above $\phi_c$. The evolution of the displacement
length scale $\ell_Q$ above $\phi_c$ in Fig.~\ref{fig:vhmax.above.below}
suggests a simple explanation for this decay. The overlap function $Q(a,\gamma)$
relax when typical displacements are of the order of $a$.  When $\phi$
increases, typical displacements in response to strain increments become
smaller, and more steps are needed to decorrelate the overlap. If these steps
are not perfectly correlated, more steps directly imply less dynamic
fluctuations and thus a reduced $\chi_4$~\cite{toni}.  This physical
explanation was made more quantitative in the empirical model of
Ref.~\cite{ballesta}, which then accounts for a nonmonotonic evolution of
$\chi_4$ across $\phi_c$.


To summarize, borrowing tools from studies of dynamic heterogeneity in
glass-formers, we provided a detailed account of the evolution of the
microscopic dynamics of an idealized model system across the jamming transition.
We found a number of original signatures, that are strikingly different from the
behaviour of viscous liquids, but might have been observed in recent
experimental reports.  Given the (relative) ease with which particle
trajectories can be monitored in suspensions, foams, or granular assemblies, we
hope our study will motivate further experimental studies along the lines of the
present study.


\acknowledgments CH acknowledges financial support from the Humboldt Foundation.

\end{document}